\documentclass[manuscript]{acmart}
\AtBeginDocument{%
  \providecommand\BibTeX{{%
    \normalfont B\kern-0.5em{\scshape i\kern-0.25em b}\kern-0.8em\TeX}}}

\setcopyright{acmcopyright}
\copyrightyear{2022}
\acmYear{2022}
\acmDOI{XXXXXXX.XXXXXXX}

\acmConference[ICMI'22]{Make sure to enter the correct
  conference title from your rights confirmation emai}{7-11 Nov}{ICMI 2022, Virtual}
\acmBooktitle{Woodstock '18: ACM Symposium on Neural Gaze Detection,
 June 03--05, 2022, Woodstock, NY} 
\acmPrice{15.00}
\acmISBN{978-1-4503-XXXX-X/18/06}

\begin{document}

\title[A Centennial Vision of Interplanetary Virtual Spaces in Turn-based Metaverse]{Beyond the Blue Sky of Multimodal Interaction: A Centennial Vision of Interplanetary Virtual Spaces in Turn-based Metaverse}

\author{Lik-Hang Lee}
\authornote{This is the corresponding author: likhang.lee@kaist.ac.kr}
\affiliation{%
  \institution{KAIST}
  \city{Daejeon}
  \country{South Korea}}
\email{likhang.lee@kaist.ac.kr}

\author{Carlos Bermejo Fernandez}
\authornote{Both authors contributed equally to this research (co-second authors).}
\author{Ahmad Alhilal}
\authornotemark[1]
\affiliation{%
  \institution{Hong Kong University of Science and Technology}
  \city{Hong Kong}
  \country{Hong Kong}
}

\author{Tristan Braud}
\affiliation{%
  \institution{Hong Kong University of Science and Technology}
  \city{Hong Kong}
  \country{Hong Kong}}

\author{Simo Hosio}
\affiliation{%
  \institution{University of Oulu}
  \city{Oulu}
  \country{Finland}
}

\author{Pan Hui}
\affiliation{%
  \institution{Hong Kong University of Science and Technology}
  \city{Hong Kong}
  \country{Hong Kong}}

\author{Esmée Henrieke Henrieke Anne de Haas}
\affiliation{%
  \institution{KAIST}
  \city{Daejeon}
  \country{South Korea}}

\renewcommand{\shortauthors}{Lee, et al.}

\begin{abstract}
Human habitation across multiple planets requires communication and social connection between planets. When the infrastructure of a deep space network becomes mature, immersive cyberspace, known as the Metaverse, can exchange diversified user data and host multitudinous virtual worlds. Nevertheless, such immersive cyberspace unavoidably encounters latency in minutes, and thus operates in a turn-taking manner. This Blue Sky paper illustrates a vision of an interplanetary Metaverse that connects Earthian and Martian users in a turn-based Metaverse. Accordingly, we briefly discuss several grand challenges to catalyze research initiatives for the `Digital Big Bang' on Mars. 
\end{abstract}





\maketitle

\section{Background}

%Facebook

During the 2020 pandemic, the implementation of many preventive measures (such as community lockdowns) changed people's daily habits and lifestyles. 
Such a sudden change has brought our way of life into a post-covid-19 era, and several aspects of our lives are gradually moving towards the Internet and virtualized platforms. %, and many old customs are being subverted. 
For instance, schools rapidly adopted the online conference platform ZOOM to conduct online classes - in many cases for a full academic year or more. Many employers and employees now not only accept but expect work-from-home.
Perhaps, the pandemic since 2020 can be regarded as one of the largest ``experiments" in history -- do people accept the movement of various functions of life into the online virtual world (i.e., the Metaverse)? 
Although we have not yet come up with a definite answer, various indications shed light on that we are open to numerous opportunities in virtual worlds, and, remarkably, irreversible changes have taken place in the metaverse era. 

The Metaverse was first mentioned by Neal Stephenson in a sci-fiction novel entitled \textit{Snow Crash}. Nowadays, the Metaverse refers to an immersive Internet, characterized by an endless and gigantic virtual environment that is able to accommodate a million users for activities (e.g., content creation) simultaneously~\cite{Lee2021WhenCM}. Since 2021, numerous virtual worlds have advent, which connect people via their avatars (e.g., \textit{Meta Horizon Workrooms~\footnote{\url{https://www.oculus.com/workrooms/}}}), although Xu et al. pinpointed the lack of consensus on the Metaverse conceptualization among research communities~\cite{chi22-metaverse}. 
An article~\cite{Lee2021AllON} recently presented a comprehensive survey regarding the technological enablers and ecosystem issues for implementing the Metaverse, in which the authors mentioned the Metaverse progression could be divided into three stages, namely (1) digital twins, (2) digital natives, and eventually (3) the convergence of physical and virtual environments. It is important to note that the three-stage progression will take at least two to three decades. 

\begin{figure}
  \includegraphics[width=0.75\columnwidth]{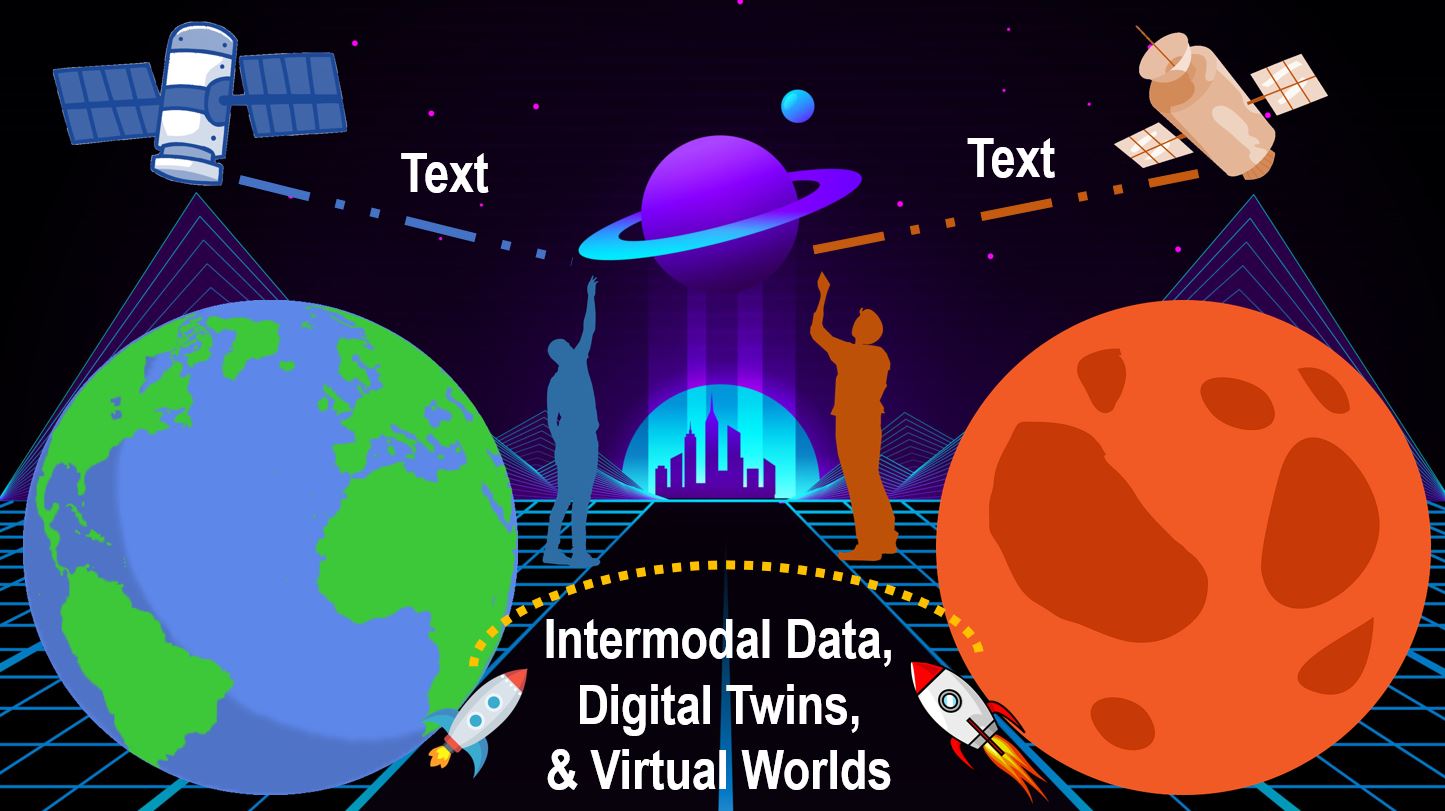}
  \caption{A vision of interplanetary Metaverse connecting Earthian and Martian: the virtual space, as a `purple planet', serves as a turn-taking communication platform that leverages multimodal approaches to realize meeting and gathering, albeit people on the Earth and Mars are apart by an ultra-long distance.}
  \label{fig:teaser}
\end{figure}

Coincidentally, the colonization of Mars is expected to achieve the milestone of establishing \textit{the first Mars city, named Nüwa}\footnote{\url{https://www.popularmechanics.com/science/a35915975/mars-city-nuwa-plans/}} by 2050. Nonetheless, most likely the first trip to Mars is a one-way trip due to the unavailability of technology for a return trip. Also, the minimum distance between the Earth and Mars (e.g., 55 million kilometers away) leads to no less than a two-year travel time with our current spaceship technology. Theoretically speaking, humans on Mars are analogous to a `community lockdown' situation on a planet-wide scale. Thus, facilitating effective communication channels between people on the two planets, and offering a way of social gathering for them, are yet to be solved. Virtual environments under the interplanetary scenario are regarded as one of the unexplored solutions driven by the Metaverse (Figure~\ref{fig:teaser}).

This paper introduces a new perspective of multimodal interaction when humans are already moving ahead to another planet. Additionally, we outline the potential roles of multimodal interaction with the metaverse and briefly present the grand challenges that respond to the rising interest in space-oriented human-computer interaction~\cite{space-chi-2021}.

\section{Interplanetary Metaverse} 
This section first explains the physical constraints of implementing an interplanetary Metaverse. Accordingly, we discuss possibilities for the people's communication driven by multimodal interaction in virtual Earthian-Martian environments. 

\paragraph{\textbf{The Primary Constraint: Martian Bandwidth.}}
The distance between the Earth and Mars is not a fixed value. It depends on the trajectory and relative orbital positions of the two planets. The varying relative positions can result in a minimum latency of around four minutes in each network utility, known as a `\textit{ping}'. In addition, the latency can go up to 20 minutes with our current IP Network (e.g., direct links), and a round-trip communication between two planets will need roughly 40 minutes\footnote{\url{https://www.forbes.com/sites/jamiecartereurope/2020/08/02/why-dont-we-have-live-video-from-mars-nasas-jaw-dropping-plans-for-laser-tv-from-the-red-planet/?sh=b3127fe18716}}. In the worst scenario, the communication breakdown occurs about every two years, when the Earth and Mars wind up on opposite sides of the Sun. %is located between Mars and the Earth. 

Furthermore, the existence of an interplanetary network can only improve the reliability of the network, also known as Delay/Disruption Tolerant Networking (DTN), but not alter the physical constraints, such as distance and the speed of light~\cite{alhilal2021roadmap}. That is, a laser link in a straight line from Earth to Mars can achieve the theoretical minimum latency of 4 minutes. Nonetheless, the direct and seamless transmission of messages and communication will not hold in practice as the Mars orbiter (i.e., a key satellite for relaying communications) is only visible to the Earth counterpart for a short time window. 
With such stringent constraints in the Martian bandwidth, NASA's Perseverence rover makes only two sessions of 15-minute communication every day. As such, high-resolution and colored still images will take several hours in transmission, and a large-size file will be divided and delivered into numerous smaller pieces. 

\begin{figure}
  \includegraphics[width=.85\columnwidth]{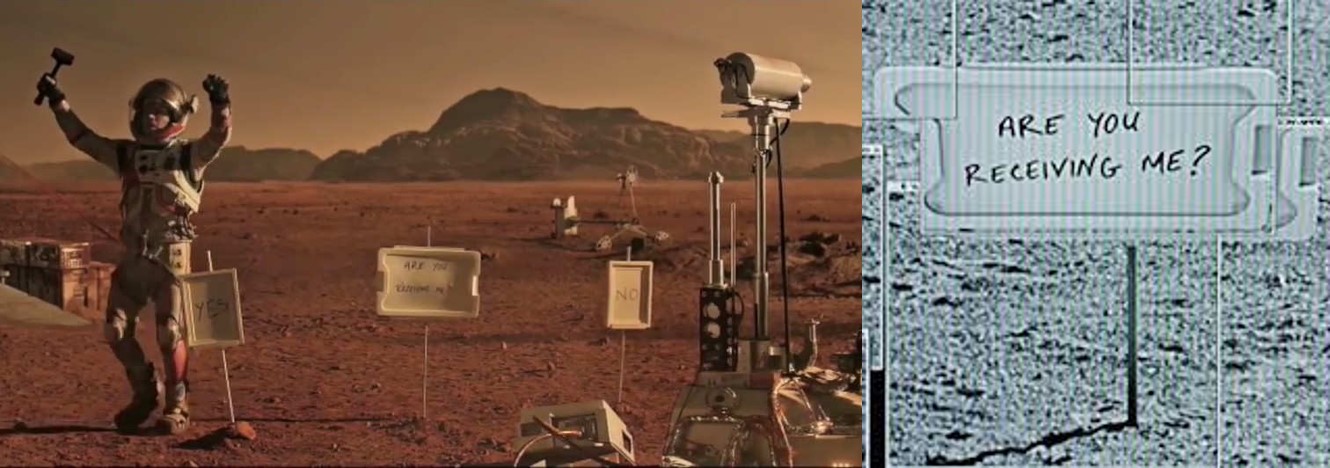}
  \caption{In the absence of other means of communication, the protagonist of Andy Weir's ``The Martian" (2011/2015) resorts to text and pictorial interaction between Mars and the Earth}
%  \caption{This illustration of Mar-Earth communication is retrieved from \textit{Andy Weir}’s sci-fi movie named The Martian. }
  \label{fig:mars-communication}
\end{figure}

\paragraph{\textbf{Possibilities of Earthian-Martian Interaction.}} At first glance, the distance and network bandwidth further make multimodal interaction challenging. Considering the constraints as mentioned above, implementing interactive virtual environments and real-time communication becomes technically infeasible. As a round-trip communication session can vary from hours to days, jitter and latency will deteriorate the user experience significantly, primarily related to the sense of presence and realism~\cite{iss19-presence-vr-jitter}. 
In other words, the most common enriched communication mediums available on the Earth, including real-time video conferencing, videos, and even colored images, are no longer applicable, not to mention the multimodal interaction in virtual worlds because of the demanding transmission of 3D graphics regarding virtual scenes and avatars. 
In the sci-fi novel and movie \textit{The Martian} (Figure~\ref{fig:mars-communication}), the Earthian-Martian communication leveraged text-based interaction and signage through black-and-white images in a turn-taking approach. Similarly, people's communication can be characterized by some type of turn-taking and non-real-time interaction, e.g., ARCAXER \footnote{\url{https://www.youtube.com/watch?v=LYVfmiIKMX0}}. As such, we have to consider these characteristics as the core consideration of designing the interplanetary Metaverse. 

\paragraph{\textbf{Values of Multimodal Interaction.}} It is worth mentioning that turn-taking interaction does not mean second-rated and is also exclusive to multimodal interaction. In fact, our daily communication occurs in a two-way conversation when one person listens while the other person speaks, or vice versa. Apart from verbal communication, people rely on other non-verbal cues, including attention indication and direction, approval, social grooming, social disruption, and interpersonal provocation~\cite{cscw-non-visual}. Therefore, multimodal interaction can serve as a strategy to express the non-verbal cues, through employing various input sensors to capture our speeches, emotions, gestures, etc., and meanwhile delivering diversified (output) cues that stimulate our five senses, commonly recognized as visual, audio, and haptic feedback. 

\paragraph{\textbf{The Avatar Turn-taking in a Virtual `Metaverse' Planet.}} 
The Metaverse acts as connection dots between multimodal enriched interaction and Earthian-Martian non-real-time turn-takings. The key idea is that, when the AI-driven embodied agents meet the Metaverse, the avatars, perhaps assisted by AIs or intelligent agents~\cite{Serenko2007EnduserAO}, can interact with other avatars in such an interplanetary virtual environment. In such context, the captured data from the user's facial expression, physical body movements, and other internal states (e.g., heartbeat rates) can be converted into a dataset representing the user interaction trace mapping to conversation dialogues. The avatars, as a type of digital twins, can leverage the dataset to reconstruct and manage high-resolution user interaction in the Metaverse, including verbal communication (e.g., text) and non-verbal interaction. 

More importantly, we have to strike a balance between rationale resource allocation and user experience (Figure~\ref{fig:teaser}). On the one hand, the bulky data, including the digital twins of the respective planet(s), avatars, and virtual scenes, will be first conveyed by physical means, i.e., rockets. As replenishment rockets will arrive on Mars regularly, the metaverse and avatars will probably get updated patches quarterly. Also, once the infrastructure on Mars becomes more mature, smaller rockets, if necessary, can deliver bulky data from Mars to the Earth.
On the other hand, the interplanetary network will primarily be responsible for transmitting text-based messages that will drive the behaviors of the embodied agents (i.e., the avatars). The text-driven turn-taking can thus maintain low-cost triggers of two-way user interaction between Earthian and Martian, reserving %. Meanwhile, AI-driven avatar behaviors can reserve 
reasonable levels of interactivity and expressiveness. % characterized by the Metaverse even non-real-time turn-taking occurs.  

%In the next section, we further detail the multi-modal interaction, including the data and technological requirements, under the three-stage progression of the Metaverse~\cite{Lee2021AllON}. 

%This year’s conference theme: Embodied Conversational Agents (ECAs) enable natural Human Computer Interaction, inspired by human-human communication. With rapid advances in multimodal analysis, dialog and synthesis technologies, intelligent ECAs are set to enter real world applications. The expected intelligence includes cognitive, social and emotional facets that humans routinely display in conversations. The theme for ICMI 2022 will revolve around making the ECAs more robust, responsible and multilingual. As such, this year, ICMI welcomes contributions on our theme for ”Intelligent and responsible Embodied Conversational Agents (ECAs) in the multilingual real world”.

%https://www.quora.com/How-will-we-support-low-latency-interplanetary-gaming-when-we-settle-on-Mars 
%https://www.msn.com/en-us/money/other/nasa-scientists-achieve-long-distance-quantum-teleportation/ar-BB1c8GyK

\section{Enabling Technology}
\begin{figure}
  \includegraphics[width=\columnwidth]{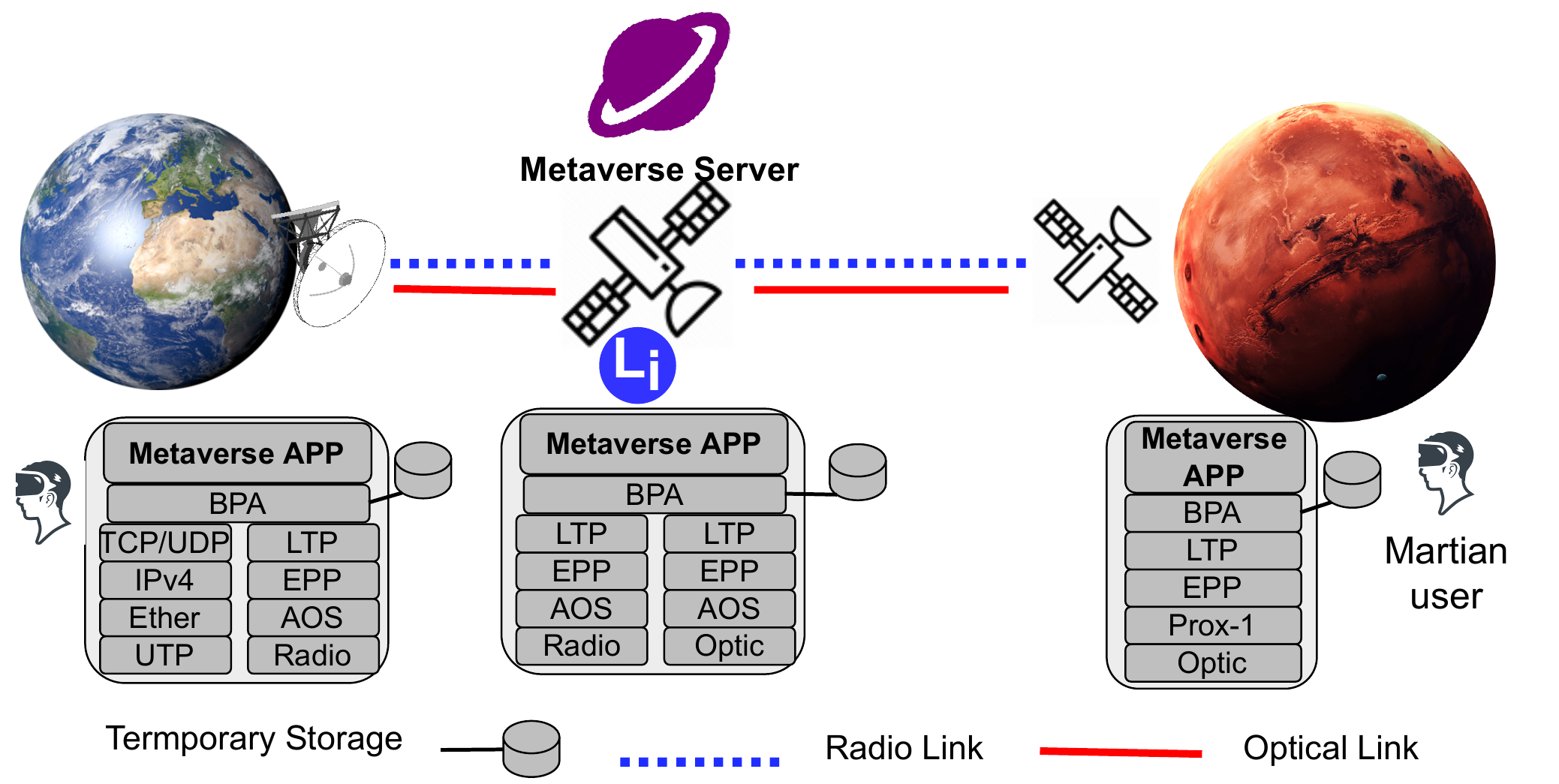}
  \caption{Underlaying communications and protocol stack to enable IPN Metaverse~\cite{alhilal2021roadmap}.}
  \label{fig:enabling-tech}
\end{figure}

Figure~\ref{fig:enabling-tech} illustrates the underlying technology to enable the interplanetary network (IPN) Metaverse. Earthians interact with Martians through an intermediary spacecraft that operates as a Metaverse server to drop the delay. The terrestrial deep space network (DSN) intercommunicates with the spacecraft using radio waves (dotted blue line) to countermeasure atmospheric absorption. Martian orbiter intercommunicates with the spacecraft radio waves or optical links (solid red lines). The spacecraft could be any spacecraft stationing at one of the Lagrangian points (L2,L3,L4,L5) depending on the relative location of the Earth to Mars. The virtual space (purple planet) is installed on the Metaverse server. The Metaverse application runs on top of a bundle protocol agent (BPA) to ensure reliable delivery as a crucial component of the end-to-end DTN architecture. 
For space links, DTN can utilize Ku-Band and Ka-band of radio waves. However, future space missions are envisioned to use the visible light spectrum to achieve much higher data rate~\cite{nasa2019optical}. This enables bandwidth-demanding multimodal data such as audio and video.

\section{Multimodal Interaction by Stages}
In this section, we further detail the roles of multimodal interaction, including the data and technological requirements, under the three-stage progression of the Metaverse~\cite{Lee2021AllON}. The embodied agents leverage the data and technology to offer functions, services, user connections, and social experiences. 
%\textcolor{red}{Figure~\ref{fig:stages} gives an overview of multimodal interaction in the consecutive stages. } 

%First of all, 
\textbf{Digital Twins} (Stage I) are digital models that duplicate the characteristics and behaviors of their physical counterparts. These high-quality models require huge amounts of data, for example to represent the properties (e.g., temperature, humidity) of the physical twins~\cite{Lee2021AllON}. At first, the information about the digital twins will be transmitted using physical storage that is delivered using the same replenishment rockets. The conveyed data will continue to update the original digital twins via periodic patches. At the same time, inspired by %and following 
the proposed roadmap in~\cite{alhilal2021roadmap}, %there will be a 
continuous deployment of relay satellites %components (e.g., satellites) to 
can enable a DTN communication system. The DTN will enable the transmission of a huge amount of information that does not require real-time communications, such as updates of existing digital twins and small quantity of new digital twins. 
Similarly, these DTN architecture will allow bidirectional transmission, overcoming the limits of sending resource-demanding rockets from Mars to the Earth. %that might be limited in the case Mars to Earth communications using rockets due to lack of resources in Mars to periodically send rockets back to earth. 

Due to the bandwidth limitations of interplanetary communications~\cite{alhilal2021roadmap}, AI-based behaviour modeling will be the core of these digital twins, regardless of humans (e.g., avatars representing a friend or family members on the Earth) or non-human (e.g., buildings or attraction sites on the Earth). The AI models behind the virtual entities can %, should include AI-based models that can 
simulate, with realism and accuracy, the properties and behaviors of their physical counterparts. %twins. 
In particular, the AIs of digital twins are well-trained to present diversified behaviors once they receive text-based instructions, albeit with delays. Also, the AIs can maintain some social responses autonomously when awaits further instructions, for the sake of user interactivity and social presence. 
These models will reduce the transmission traffic of (1) sensor-related data, such as temperature and object motions, and (2) human-driven behaviors in the interplanetary network. 

%We further elaborate the necessity of digital twins and AI models in the turn-taking metaverse, with the below example. 
Following a prior work on embodied virtual agents~\cite{cassell2001embodied}, the created avatars and other virtual assets should embed smart features to facilitate the interaction (e.g., communication) between users in the interplanetary metaverse. 
As such, the presence %creation 
of avatars and their behaviors will be deteriorated by noticeable delays. The multimodal interaction, e.g., gestures and speech tones/pitches, will be generated by intelligence in the digital twin of avatar models locally on Mars. % due to the limitations of real-time communications (delays), should embed intelligence in their avatars. 
%Following a prior work on embodied virtual agents~\cite{cassell2001embodied}, the created avatars and other virtual assets should embed smart features to facilitate the interaction (e.g., communication) between users in the interplanetary metaverse. 
Moreover, we can see the creation of this embodied avatar as clones of real users (e.g., an Earthian 5.5 million away), where gestures, facial expressions, and emotions emulate the actual user. %simulate the real user. 
Therefore, we need an intelligent system that captures multimodal data, such as user movements, facial expression, speech, heartbeat rate, mental state, emotions, nerve activities of real users on a planet, to have a faithful representation of the real user as embodied avatars on another planet. These embodied virtual agents can speed up the turn-taking in the text-alike format. % of text delivery. 
In contrast, turn-taking of transmitting graphics and videos (e.g., gestures) causes more %overcome the challenges turn-taking with 
severe delays by keeping some realistic reactions for the avatars at distal, while the physical Martian user will suffer from prolonged waiting and quit the turn-taking Metaverse. 
Finally, the digital twins can connect to local output devices on Mars that enrich feedback cues to Martian users. For instance, the avatars, after receiving a message -- `\textit{Give me five}', can give haptic feedback through wearables with touch and kinesthetic features attached to the Martians' bodies. 
%of the avatars while waiting for the real responses (from the physical user). Due to the lack of multitude of people in the first Mars colonies these digital twins should also consider the representation of haptics (e.g., touch, kinesthetic) feedback using wearables attached to the martians' bodies. 

Second, once a significant size of digital twins is available, the main activities in the Metaverse will shift to the creation of digital content in the Martian environment (\textbf{Digital Natives}, Stage II). This stage will become feasible when a growing bandwidth of DSN exists. As we expect that turn-taking content creation and information exchange will become more frequent, we have to half the latency, i.e., axing each turn from four to two minutes, by deploying a Metaverse server between the Earth and Mars, denoted as the `\textit{purple planet}' in Figure~\ref{fig:enabling-tech}).  
%\paragraph{\textbf{Digital Natives.}} Once 
%the digital copies of the physical world are established, the creation of digital content in the martian environment will play a main role. 
Metaverse participants %Members of the metaverse 
(Earthians or Martians) will enable a new paradigm with embodied virtual agents %new worlds 
and turn-based interactions in the virtual counterpart, connecting to diversified %the different 
ecosystems, e.g., culture, gaming, social, and economy. The interplanetary Metaverse will require protocols for managing user instructions, and (multi-)agent behaviors. Meanwhile, the participants will work on virtual tasks and create contents in iterations. This opens research opportunities of turn-based collaboration among multiple users, and the design space of such tasks that contain virtual objects and their behaviors. 
%turn-taking content creation 
%synchronization 

%\paragraph{\textbf{Convergence between Virtual Planets and Respective Planets.}}
Next, the third stage depicts the convergence of (physical) blue, red, and (virtual) purple planets (\textbf{Convergence between Virtual Planets and Respective Planets}, Stage III). We foresee that Metaverse servers will gradually form a distributed network, potentially resulting in lower latency, improved bandwidth, and hence diversified user interaction distal. These servers, local to each planet, would be interconnected through well-established links, enabling faster and more stable transmission of content across the solar system~\cite{alhilal2021roadmap}. From the Martians' perspective, we can see the metaverse as a door to explore and travel inside %visit 
the virtual earth. Digital twins that replicate buildings, e.g., museums, can be virtually explored by the Martians in the metaverse. 
Moreover, we see embodied avatars as digit twins of physical users, allowing the Martians to have quasi-real interactions (e.g., chatting) in the turn-taking scenarios. 
%robotics presence
Alternatively, both Martian and Earthian can leverage social robots to achieve telepresence in the physical means, visiting places and people remotely.  

%\paragraph{\textbf{Scenario Construction.}} 

%\newpage
\section{Conclusion and Grand Challenges}
The blue sky should not become the limit of multimodal interaction. Meanwhile, the Metaverse, initiated by the concept of digital twins, will reach interplanetary users. %should go beyond the blue sky of the Earth. Meanwhile,  
This paper proposes an interplanetary virtual space of turn-taking. %, \textit{the purple planet}, supported by turn-taking and low-cost user interaction. %, in which the key enabler is embodied virtual agents emulating user behaviors and hence intelligent responses. 
We outline several grand challenges  in such an interplanetary Metaverse, as follows.

\paragraph{\textbf{Trust and Multiple Identities.}} The use of virtual entities has an impact on the social interaction between users. The metaverse allows users to create a myriad of avatar representations, from self avatar that realistically represents the owner of the avatar to multiple identities such as animals, different gender, and impersonation of other avatars. The representation of the avatar has a strong impact on social dynamics~\cite{collingwoode2021impact,steed1999leadership}. In the case of embodied avatars, users' representation in the metaverse goes beyond visuals. Gestures, emotions, and other data that simulate the physical owner of the avatar are essential features in building the trust of other users. Reactions of these embodied avatars that are not realistic or go close to the uncanny valley can reduce users' trust in the metaverse. In this case, the more feedback and multimodal interactions, the more risk of users stopping trusting an entity when these are not realistic or do not represent the behavior and actions of the physical owner (self-representations).

\paragraph{\textbf{Integrity of Multimodal Data.}} As we have described, the embodied virtual agents will require data from sensors to simulate the physical owners' behavior. AI models will rely tightly on the quality of the sensed data to represent the physical users while waiting for users' responses. The integrity of the data involves monitoring the sensors (hardware), data collection algorithms, and AI models (training and testing). The data integrity should also be resilient to possible attacks, such as biasing the data collection process. As we have seen in prior challenges, when the embodied virtual agents do not present realistic behaviors, the users will feel uncomfortable and modify their social dynamics with such virtual agents.

\paragraph{\textbf{Resolution of the turn-taking Metaverse.}}
Interplanetary communication is unequivocally afflicted with long latency, in the order of minutes to days. Novel semi-synchronous communication modalities will need to be developed for such constraints. These modalities will rely on both explicit (enforced) and implicit (cultural) turn-taking protocols that oversee whose turn it is to communicate. These protocols already implicitly exist in the text- and voice-based instant messaging applications such as WhatsApp, and to a lesser extent in email communication. However, how to integrate such protocols within the richer spectrum of interaction modalities offered by the Metaverse, close to face-to-face communication, at a scale spreading from minutes to days between the transmission of messages remains unknown. Specifically, these protocols will aim to manage or hide the downtime caused by network transmission. AI agents may act as intermediary nodes in the communication chain, enabling more natural, albeit less direct, communication between two users on different planets while enforcing such semi-synchronous protocols. Although human-agent synchronous communication is currently the subject of significant research~\cite{10.1145/3386867}, there has been little work on human-agent-human communication combining synchronous human-agent communication with asynchronous human-human communication.

%What the avatar doing autonomously during waiting the turn? How the user understands their behaviours?
%User Expectation with AI-driven Scenes and Avatars.
%\paragraph{\textbf{Protocols for Turn-taking content creation.}}
%Imagine you and your friend trying to communicate using Whatsapp in discontinuous network connectivity! What you do is record and send him, then he replies and so on 

\paragraph{\textbf{Embodied virtual agents and Human-scale space.}} %simo
The space under the human perspective, i.e., human-scale space~\cite{simo}, will influence the user behaviors. Even in an identical (virtual) space, users with embodied virtual agents own diversified perspectives, in addition to the less favorable condition of non-real-time user interaction. Human users have built-in sensing organs that help us understand and navigate the world through the five primary senses. Remarkably, we have a lot of subconscious understandings of places. For example, a dark- and blue-colored house can be regarded as a haunted house. However, if we feed these inputs (as texts) into embodied agents, they may give a house with bad weather. Thus, this leaves questions about how to construct a human-centric perception of human-scale space in a turn-based metaverse. 
% We are moving in cities, environments.
% Interactive Services and Spaces , which deals about understanding the human scale. So most environments that we usually operate in are in the human scale, which is our eye level spaces. Space is where we interact with objects. The human scale is also very far the most common space, or scale in virtual reality and virtual environments.
% We have built in hardware sensing organs, sensing organs that help us understand, navigate the world through these five primary senses. But it goes beyond that. We built abstractions based on the input from those sensory organs. We have a lot of subconscious understandings of places that cannot be described through the input of those senses alone. So if I feed those inputs into a machine, they won't understand things like, uh, haunted house vibes. 
% What is the overall user experience of using services in a given human scale space? 

\paragraph{\textbf{Scalability of embodied agents in the Metaverse.}} %Tristan/Admad
The IPN Metaverse should scale up to adapt to the evolutionary IPN architecture. In particular, it must evolve as IPN communication architecture evolves from a Near-term (low data rate and non-network) to a Long-term architecture (high data rate and availability), passing through a Mid-term architecture~\cite{alhilal2021roadmap,alhilal2019sky}. Ultimately, it must cope with the bi-directional and high bandwidth-demanding multimodal interaction. 
As such, we have to prioritize the multimodal data for digital twins and user interaction in the aforementioned timeline. The availability of multimodal data will govern the behaviors of digital twins, including embodied virtual agents, and thus impact user perception of the immersive spaces, e.g., object, social and spatial presences. 
\bibliographystyle{ACM-Reference-Format}
\bibliography{output}

%\appendix









\end{document}